\documentclass[
showpacs,preprintnumbers,amsmath,amssymb]{revtex4}

\usepackage{amsmath}

\begin{document}

\tolerance=5000

\def\be{\begin{equation}}
\def\ee{\end{equation}}
\def\bea{\begin{eqnarray}}
\def\eea{\end{eqnarray}}
\def\tr{{\rm tr}\, }
\def\nn{\nonumber \\}
\def\e{{\rm e}}
\title{Scalar-Tensor theories and current Cosmology}

\author{Diego S\'{a}ez-G\'{o}mez$^{1,}$\footnote{Electronic address:saez@ieec.uab.es}}
\affiliation{$^{1}$Consejo Superior de Investigaciones Cient\'\i ficas
ICE/CSIC-IEEC, Campus UAB, Facultat de Ci\`encies, Torre
C5-Parell-2a pl, E-08193 Bellaterra (Barcelona) Spain}
\begin{abstract}
Scalar-tensor theories are studied in the context of cosmological evolution, where the expansion history of the Universe is reconstructed. It is considered quintessence/phantom models, where inflation and cosmic acceleration are reproduced. Also, the non-minimally  coupling regime between the scalar field and the Ricci scalar is studied and cosmological solutions are obtained.  The Chamaleon mechanism is shown as a solution of the local gravity tests problems presented in this kind of theories. 
\end{abstract}
\maketitle
\section{Introduction}
Since the supernovae observations were analyzed in 1998 (see \cite{DiscAcc}), the majority of the scientific community has accepted that the Universe is in an accelerated expansion phase. As it was unknown the possible mechanism to make this kind of repulsion gravity, this was called dark energy (for a review of possible candidates see \cite{DDE}), which it is supposed to have an equation of state parameter (EoS) less than $-1/3$, and an energy density close to the critical density $\rho_{DE}\sim 10^{-3}eV$. The main candidate to dark energy has been the cosmological constant, which may  represent the vacuum energy but there is unknown explanation as to why the energy density is of the order of the critical density, much smaller than vacuum energy density predicted by quantum field theory (for a review on the comological constant problem see \cite{CosmCons}). Then, some others candidates have been proposed, one of them is the quintessence/phantom scalar field models (\cite{phantom2},\cite{phantom},\cite{QuintPhamUnf}, \cite{Ex.1} and \cite{Ex.2}), where a scalar field minimally coupled is included, and the accelerated expansion is reproduces by this single field,  whose EoS parameter is around $-1$. Phantom models where the EoS parameter is less than $-1$, have not been excluding by observational data, and together with quintessence scalar field models, it constitutes a good and simple candidate to dark energy. Currently, the main purpose of scalar-tensor theories is not just to explain the cosmic acceleration but to reproduce all the expansion history from the early accelerated expansion called inflation to the cosmic acceleration. Also, scalar-tensor theories with scalar fields non minimal coupled to the Ricci scalar are currently considered, which may present problems with local gravity tests, although these are avoided by using the so-called chamaleon mechanism (see \cite{chamaleon}), such that this kind of Brans-Dicke theory are not excluded.  \\
At the present article, we review some solutions of quintessence/phantom models, where some examples are given and the evolution of the Universe is reproduced. The possible future singularities in the phantom epochs are studied. Also Brans-Dicke-like theory is considered where cosmic solutions are reconstructed, and the chamaleon mechanism is shown in such a way that the local gravity constraints may be avoided by this kind of theories.

\section{Quintessence and phantom scalar-tensor cosmology}

The Quintessence/phantom scalar field models are well studied (for a review see \cite{DDE}), they are presented as an  explanation of the current cosmological acceleration.  Also  the majority of inflation models are constructed by a single scalar field called inflaton. Recently, the main task in scalar-tensor theory is the possibility to unify both inflation and late-time acceleration by using a single scalar-field (see for example  \cite{QuintPhamUnf}, \cite{Ex.1}, \cite{Ex.2}), this real possibility is shown above by several examples. Let us start consider the action that define these kind of models, we consider a Universe filled with some kind of matter whose equation of state (EoS) is given by $p_m=w_m\rho_m$ (here $w_m$ is a constant) and a scalar field minimally coupled to gravity, the action is given by:
\begin{equation}
S=\int dx^{4}\sqrt{-g}\left[ \frac{1}{2\kappa^{2}}R
 \pm \frac{1}{2} 
\partial_{\mu} \varphi \partial^{\mu }\varphi -V(\varphi )+L_{m}\right]\ ,
\label{1.1}
\end{equation} 
here the sing ($\pm$) in front of the kinetic term define the nature of the scalar field, where the phantom-like field is given by (+) and the quintessence-like field by (-). The lagrangian $L_m$ is the matter lagrangian density. The field equations are obtained by varying the action (\ref{1.1}) with respect $g^{\mu\nu}$:  
\be
R_{\mu\nu}-\frac{1}{2}g_{\mu\nu}=\frac{\kappa^2}{2}\left[ {2} T^{(m)}_{\mu\nu}+T^{(\varphi)}_{\mu\nu}\right] \ ,
\label{1.2}
\ee
where $T^{(m)\mu}_{\nu}=(\begin{array}{cccc}-\rho_m & p_m & p_m &p_m) \end{array}$ and $T^{(\varphi)\mu}_{\nu}=\partial^{\mu}\varphi\partial_{\nu}\varphi-g^{\mu}_{\nu}\left(  \frac{1}{2}g^{\alpha\beta}\partial_{\alpha}\varphi\partial_{\beta}\varphi+V(\varphi)\right) $. We assume a flat FRW spacetime, the metric is given by:
\begin{equation}
ds^{2}=-dt^{2}+a^2(t) \sum_{i=1}^{3} dx_{i}^{2}\ .
\label{1.3}
\end{equation}  
Then, the corresponding Friedmann equations are written as:
\be
H^{2} = \frac{\kappa ^{2}}{3}\left( \rho _{m}
+\rho_{\varphi}\right)\ , \quad \quad \dot H = -\frac{\kappa ^{2}}{2}\left(
\rho _{m}+p_{m}+\rho _{\varphi }+p_{\varphi }\right)\ ,
\label{1.4}
\ee
where $\rho_{\phi}$ and $p_{\phi}$ are given by:
\be
\rho _{\varphi } = \mp\frac{1}{2} \, {\dot \varphi}^{2}
+V(\varphi)\ ,\quad \quad
p_{\varphi } = \mp\frac{1}{2} \, {\dot \varphi}^{2}-V(\varphi)\ ,
\label{1.5}
\ee
here the phantom case is given by the case of negative value of the kinetic term, just as it is the definition for a phantom field, where the weak energy condition ($\rho+p\geq0$) is violated. In other hand, by varing the action (\ref{1.1}) with respect $\phi$, the scalar field equation is obtained:
\be
\ddot{\varphi}+3H\dot{\varphi}\mp V'(\varphi)=0\ ,
\label{1.6}
\ee
here the prime on the potential term means a derivative respect $\phi$. Then, by the equations (\ref{1.4}) and (\ref{1.6}) the  solution for the scale parameter may be obtained. In addtion, the continuity equation for the matter term may be useful to find out the solution:
\be
\dot{\rho_m}+3H(\rho_m+p_m)=0\ .
\label{1.7}
\ee
Then, the matter component, which is suposed to have an EoS given by $p_m=w_m\rho_m$, behaves as $\rho_m\varpropto a^{-3(1+w_m)t}$. A simple way to resolve the above equations may be done by redefining the scalar field as:
\be
\varphi=\int^{\phi}d\phi\sqrt{\pm\omega(\phi)}\ ,
\label{1.8}
\ee  
where the sign depends on the sign of $\omega(\phi)$, i.e. on the phantom behaviour of the scalar field. Hence, the action (\ref{1.1}) takes the form:
\begin{equation}
S=\int dx^{4}\sqrt{-g}\left[ \frac{1}{2\kappa^{2}}R
 - \frac{1}{2} \omega (\phi)
\partial_{\mu} \phi \partial^{\mu }\phi -V(\phi )+L_{m}\right]\ ,
\label{1.9}
\end{equation}
and the equations (\ref{1.5}) are written as:
\be
\rho _{\phi } = \frac{1}{2} \omega (\phi )\, {\dot \phi}^{2}
+V(\phi)\ ,\quad \quad
p_{\phi } = \frac{1}{2} \omega (\phi ) \, {\dot \phi}^{2}-V(\phi)\ .
\label{1.10}
\ee 
Then, combining the Friedmann equations (\ref{1.4}) with (\ref{1.10}), one obtains:
\be
\omega (\phi ) \, \dot{\phi ^{2}}
= -\frac{2}{\kappa^{2}}\dot{H}-(\rho _{m}+p_{m})\ ,\quad \quad
V(\phi ) = \frac{1}{\kappa ^{2}}
\left( {3H}^{2}+\dot{H} \right) -\frac{\rho_{m}-p_{m}}{2}\ .
\label{1.11}
\ee
Now we can use the continuity equation for the matter contribution (\ref{1.7}), and $V(\phi)$ and $\omega( \phi)$ can be expressed as:
\be
\omega (\phi ) = -\frac{2}{\kappa ^{2}}f^{\prime }(\phi )
 -(w_{m}+1)F_{0} \e^{-3(1+w_{m})F(\phi )}\ ,\quad
V(\phi ) = \frac{1}{\kappa ^{2}}
\left[ {3f(\phi)}^{2}+f^{\prime }(\phi ) \right]
+\frac{w_{m}-1}{2}F_{0}\, \e^{-3(1+w_{m})F(\phi )}\ ,
\label{1.12}
\ee
where $f(\phi)\equiv F'(\phi)$, $F$ is an arbitrary differentiable function of $\phi$, and $F_0$ is an integration constant. Then the following solution is found (see \cite{QuintPhamUnf}):
\be
\phi=t \quad, \qquad H(t)=f(t)\ ,
\label{1.13}
\ee
which leads to
\begin{equation}
a(t)=a_{0}\e^{F(t)}, \qquad a_{0}=\left(
\frac{\rho _{m0}}{F_{0}}\right) ^{\frac{1}{3(1+w_{m})}}.
\label{1.14}
\end{equation}
Then, by the formulation shown above, one may construct models of the Universe in the mark of scalar-tensor theories where the cosmic acceleration and also inflation are well reproduced. A useful parameter to study the evolution of these models is defined by:
\be
w_{eff}=\frac{p}{\rho�}=-1-\frac{2\dot{H}}{3H^2�}\ ,
\label{1.14a}
\ee
where,
\be
\rho=\rho_m+\rho_{\phi}\quad , \qquad p=p_m+p_\phi
\label{1.14b}
\ee
In order to study the accelerated epochs of the Universe, it is convenient to write also the expression for the acceleration of the scale parameter:
\be
\frac{\ddot{a}}{a×}=H^2+\dot{H}=-\frac{\kappa^2}{6×}(\rho+3p)\ .
\label{1.14c}
\ee
Then, the Universe will be in an accelerated phase when the effective fluid is given by $p<-1/3\rho$. Some examples are presented below, where the above formulation is used to reconstruct the evolution of the Universe. Note that in spite of the simplicity of the formulation derived, in general the expressions for the scalar potential and the kinetic term will be very complex, even for simple solutions of the Hubble parameter, as it will be seen in the following examples.
\subsection{Inflation}
As a first example, we show a simple model of inflation. The majority of inflation models are given by a single scalar field, where the lagrangian for the scalar field is $L=-(1/2)(\partial\phi)^2-V(\phi)$.  Some conditions on the kinetic term and on the scalar potential are impossed in order that inflation could take place (see \cite{Cosmology}). These conditions called slow-roll conditions establishes that $\ddot{\phi}<<3H\dot{\phi}$ and $\dot{\phi}^2<<V(\phi)$, and they may be cast in a more usseful forms, the so called slow-roll parameters, which are given by: 
\be
\epsilon=\frac{1}{3\kappa ^{2}}\left( \frac{V'}{V}\right)^2  \ll 1\ ,\quad
\eta= \frac{1}{3\kappa ^{2}}\frac{V''}{V} \ll 1 \ .
\label{In.1}
\ee
Then, the Friedmann Equations and the equation of motion for the scalar field (\ref{1.6}) are written as:
\be
H^2\approx\frac{\kappa^2}{3×}V(\phi)\ , \qquad 3H\dot{\phi}+V'(\phi)\approx0\ ,
\label{In.2}
\ee
hence, by an specific potential, the solution for the equations (\ref{In.2}) is found. The inflation period is  characterized by an accelerated expansion in order to explain the horizon problem or the flatness problem (see \cite{Cosmology}), then we may choose a potential that holds the conditions (\ref{In.1}) and produces an accelerated expansion. As a simple and very well known inflation model, we choose the following scalar potential:
\be
V(\phi)=V_0\exp{\sqrt{\frac{{2\kappa^2}}{�\alpha}}\phi}\ ,
\label{I.3}
\ee
which gives a solution of the type $a(t)\sim t^{\alpha}$, where for $\alpha>1$ the accelerated expansion takes place. Then, by this very simple example, it is shown how an scalar field reproduces the inflationary period of the Universe. In the following examples, some models are reconstructed where the cosmic acceleration takes place, and even some models  where a unified scenario of the inflation and cosmic acceleration is presented.

\subsection{Cosmic Acceleration}

Let us now analize an example of how an scalar field may reproduce the behaviour of the dark energy in a natural way. By the action given in (\ref{1.9}),  the reconstruction of the kinetic and potential term for the scalar field given in equations (\ref{1.12}) and (\ref{1.13}) is used. In this case, we consider  the function $f(\phi)$  studied in Ref.\cite{Ex.1}, and which is given by:
\be
f(\phi)=H_0+\frac{H_1}{\phi�}\ ,
\label{Cos.1}
\ee
where $H_0$ and $H_1$ are positive constants. Then, by equation (\ref{1.12}), the expressions for the kinetic term and the scalar potential are given by:
\bea
\omega(\phi)=\frac{1}{\kappa^2�}\frac{2H_1}{\phi^2�}- (w_m+1)F_0\phi^{-3H_1(1+w_m)}\e^{-3H_0(1+w_m)\phi}\ , \nn
V(\phi)=\frac{1}{\kappa^2�}\left[\frac{H_1(3H_1-1)}{\phi^2�}+\frac{2H_0H_1}{\phi�}+H_0 \right]+\frac{w_m-1}{2�}F_0\phi^{-3H_1(1+w_m)}\e^{-3H_0(1+w_m)\phi}\ .
\label{Cos.2}
\eea
Then, by (\ref{1.13}) the solution for the Hubble parameter is found:
\be
H(t)=H_0+\frac{H_1}{t�}\ , \qquad a(t)=a_0t^{H_1}\e^{H_0t}\ .
\label{Cos.3}
\ee
Note that the solution (\ref{Cos.3}) behaves as an effective cosmological constant at late times. For small times and a choice given by $w_m=0$ (dust matter) and $H_1=2/3$, the Universe is dominated by matter at early times. This may be seen clearer by studying the effective parameter (\ref{1.14a}):
\be
w_{eff}=-1+\frac{2H_1}{H^2_1+H^2_0\phi^2+2H_0H_1\phi�}\ .
\label{Cos.4}
\ee
Then, the scalar field characterized by the kinetic and potential terms (\ref{Cos.2}) reproduces  a Universe where the dust matter dominates at the begining when $t\rightarrow0 $ and the effective parameter of the EoS(\ref{Cos.4}) $w_{eff}\rightarrow 0$. At late times, the scalar field begins to dominate, and to produce an accelerated expansion $w_{eff}\rightarrow-1$ ($t\rightarrow \infty$), similar to an effective comological constant. Note that in this case, the energy density of the dark energy and  its EoS parameter, is not constant and change with time. This fact, comoon in all quintessence/phantom models, opens the possibility to compare them with the cosmological constant model, and then to establish the evolution of the EoS by the observational data to constraint the models.

 \subsection{Inflation and late-time acceleration unified}

Let us now consider some models where the inflation and late-time acceleration period are reproduced by the same scalar field. In this case, it is considered an scalar field that passes through quintessence and phantom epochs, and the possible future singularities (for a classification of future singularities, see \cite{FutSing}) driven by the phantom behaviour are studied. By the reconstruction (\ref{1.12}) and (\ref{1.13}), we consider as a first example the choice (see \cite{Ex.2}):
\begin{equation}
f(\phi)=\frac{H_0}{t_s-\phi}+\frac{H_1}{\phi^2}\ .
\label{1.18}
\end{equation}
We take $H_0$ and $H_1$ to be  constants and $t_s$ as the Rip
time, as specified
below. Using (\ref{1.12}), we find that the kinetic
function and the scalar potential are
\bea
\omega(\phi) &=& -\frac{2}{\kappa^2}\left[
\frac{H_0}{(t_s-\phi)^2}-\frac{2H_1}{\phi^2}
\right] -(w_m+1)F_0\left(
t_s-\phi\right)^{3(1+w_m)H_0}\exp\left[
\frac{3(1+ w_m)H_1}{\phi}\right]\ , \nn
&&\nonumber \\
V(\phi) &=& \frac{1}{\kappa^2}\left[
\frac{H_0(3H_0+1)}{(t_s-\phi)^2}
+\frac{H_1}{\phi^3}\left(\frac{H_1}{\phi}-2 \right)\right]
+\frac{w_m-1}{2}
F_0\left(t_s-\phi \right)^{3(1+ w_m)H_0}
\e^{\frac{3(1+ w_m)H_1}{\phi}}\ ,
\label{1.19}
\eea
respectively. Then, through the solution~(\ref{1.13}), we
obtain the Hubble parameter and the scale factor
\be
H(t)=\frac{H_0}{t_s-t}+\frac{H_1}{t^2}\ ,\quad
a(t)=a_0\left( t_s-t\right)^{-H_0} \e^{-\frac{H_1}{t}}\ .
\label{1.20}
\ee
Since $a(t)\to 0^{+} $ for $t\to 0$, we can fix $t=0$ as the
beginning of the universe. On the other hand, at $t=t_s$ the
universe reaches a
Big Rip singularity, thus we keep $t<t_s$. In order to study the different stages that our
model will pass through, we calculate the acceleration parameter and the first
derivative of the Hubble parameter. They are
\be
\dot H=\frac{H_0}{(t_s-t)^2}-\frac{2H_1}{t^3}\ ,\quad
\frac{\ddot a}{a}=H^2+\overset{.}{H}=\frac{H_0}{(t_s-t)^2}(H_0+1)
+ \frac{H_1}{t^2}\left(\frac{H_1}{t^2}-\frac{2H_1}{t}
+\frac{2H_0}{t_s-t}
\right)\ .
\label{1.21}
\ee
As we can observe, for $t$ close to zero, $\ddot a/a>0$, so that
the universe is accelerated during some time. Although this is
not a phantom
epoch, since $\dot H<0$, such stage can be interpreted as
corresponding to the
beginning of inflation. For $t>1/2$ but $t\ll t_s$, the universe
is in a
decelerated epoch ($\ddot a/a<0$). Finally, for $t$ close to $t_s$,
it turns out that $\dot H>0$, and then the universe is
superaccelerated, such acceleration being of phantom nature and
ending in a Big Rip singularity at $t=t_s$, where the scale parameter $a(t)\rightarrow\infty$, and the Ricci scalar $R=6(2H^2+\dot{H})\rightarrow\infty$.  \\

Our second example also exhibits unified inflation and
late time
acceleration, but in this case we avoid phantom phases and, therefore, Big Rip
singularities. We consider the following model given in Ref.\cite{Ex.2}:
\begin{equation}
f(\phi )=H_{0}+\frac{H_{1}}{\phi ^{n}}\ , \label{1.22}
\end{equation}
where $H_{0}$ and $H_{1}>0$ are constants and $n$ is a positive
integer (also constant). The case $n=1$ yields an initially
decelerated universe and a late time acceleration phase. We
concentrate on cases corresponding to $n>1$ which gives, in
general, three epochs: one of early acceleration (interpreted as
inflation), a second decelerated phase and,
finally, accelerated expansion at late times. In this model, the
scalar potential and the kinetic parameter are given, upon
use of Eqs.~(\ref{1.12}) and (\ref{1.22}), by
\bea
\omega (\phi ) &=&\frac{2}{\kappa ^{2}}\frac{nH_{1}}
{\phi^{n+1}}
 -(w_{m}+1)F_{0} \, \e^{-3(w_{m}+1)\left( H_{0}\phi
 -\frac{H_{1}}{(n-1)
\phi^{n-1}}\right) }, \label{1.23} \\
&&\nonumber\\
V(\phi ) &=&\frac{1}{\kappa ^{2}}\frac{3}{\phi ^{n+1}}\left[
\frac{\left(
H_{0}\phi ^{n/2}+H_{1}\right)^{2}}{\phi ^{n-1}}-\frac{nH_{1}}{3}\right]
+\frac{w_{m}-1}{2}F_{0} \, \e^{-3(w_{m}+1)\left( H_{0}\phi -
\frac{ H_{1}}{(n-1)\phi ^{n-1}}\right) }.
\eea
Then, the Hubble parameter given by the solution (\ref{1.13})
can be written as
\be
H(t) = H_{0}+\frac{H_{1}}{t^{n}}\ , \label{1.24} \quad
a(t) = a_{0}\exp \left[
H_{0}t -\frac{H_{1}}{(n-1)t^{n-1}}\right]\ .
\ee
We can fix $t=0$ as the beginning of the universe because at
this point $a\to 0$, so $t>0$. The effective EoS
parameter~(\ref{1.14a}) is
\begin{equation}
w_{\rm eff}=-1+\frac{2nH_{1}t^{n-1}}{\left(
H_{0}t^{n}+H_{1}\right) ^{2}}\ .
\label{1.25}
\end{equation}
Thus, when $t\to 0$ then $ w_{\rm eff}\to -1$ and we
have an acceleration epoch, while for $t\to \infty $,
$w_{\rm eff} \to -1$ which
can be interpreted as late time
acceleration. To find the phases of acceleration and deceleration for $t>0$,
we study $\ddot a/a$, given by:
\begin{equation}
\frac{\ddot a}{a}=\dot H +H^{2}=-\frac{nH_{1}}{t^{n+1}}
+\left( H_{0}+\frac{H_{1}}{t^{n}}\right) ^{2}\ . \label{1.26}
\end{equation}
For sufficiently large values of $n$
we can find two positive zeros of this function, which means
two corresponding
phase transitions. They happen, approximately, at
\begin{equation}
t_{\pm }\approx \left[ \sqrt{nH_{1}} \,\, \frac{\left( 1\pm
\sqrt{1-\frac{4H_{0} }{n}} \, \right)}{2H_{0}} \, \right]^{2/n}\ ,
\label{1.27}
\end{equation}
so that, for $0<t<t_{-}$, the universe is in an accelerated
phase interpreted as an inflationary epoch; for $t_{-}<t<t_{+}$
it is in a decelerated phase (matter/radiation dominated); and,
finally, for $t>t_{+}$ one obtains late time acceleration, which
is in agreement with the current cosmic expansion.

\section{Brans-Dicke-like Cosmology}

Let us now consider an scalar-tensor theory where the scalar field $\phi$ does not couple minimally to Ricci scalar in the action for the gravity field (\ref{1.1}). This kind of theories were suggested in 1961 by Brans and Dicke (see \cite{Brans-Dicke}) in the context to include Mach's principle in a theory of gravity that in some limit  General Relativity(GR) was recovered, such that  GR was constrasted. Recently, this kind of theories  have become important in the context of Cosmology in order to explain the current acceleration of the Universe and even the inflation too. Also, the mathematical equivalence to some modified gravity theories as F(R)-theories, make them to play an important role (see \cite{F(R)}). In other hand, the main problems of these theories, as the coupling appeared with the normal matter when  a conformal transformation is performed, it seems to be resolved by the so-called chamaleon mechanism, which avoid possible violations of the Equivalence principle at small scales.

\subsection{Reconstruction of comological solutions}   

In the preceding section we have considered an action,
(\ref{1.1}), in which the
scalar field is minimally coupled to gravity. In the present
section, the scalar
field couples to gravity through the Ricci scalar
(see \cite{think} for a review on cosmological
applications). We begin from the action
\begin{equation}
S=\int d^{4}x\sqrt{-g} \left[(1+f(\phi))\frac{R}
{\kappa^{2}}-\frac{1}{2} \omega (\phi ) \partial_{\mu }\phi
\partial^{\mu }\phi -V(\phi )\right]\ ,
\label{2.1}
\end{equation}
where $f(\phi )$ is an arbitrary function of the scalar field
$\phi $. Then,
the effective gravitational coupling depends on $\phi $, as
$\kappa_{eff}=\kappa [1+f(\phi)]^{-1/2}$. One can work in the
Einstein frame,
by performing the scale transformation
\begin{equation}
g_{\mu \nu }=[1+f(\phi )]^{-1} \widetilde{g}_{\mu \nu }\ .
\label{2.2}
\end{equation}
The tilde over $g$ denotes an Einstein
frame quantity. Thus, the action~(\ref{2.1}) in such a frame
assumes the form \cite{CT}
\begin{equation}
S=\int d^{4}x\sqrt{-\widetilde{g}}\left\{
\frac{\widetilde{R}}{2 \kappa ^{2}}
 -\left[ \frac{ \omega (\phi )}{2(1+f(\phi))}+\frac{6}
{\kappa^{2}(1+f(\phi ))}
\left( \frac{ d(1+f(\phi)^{1/2})}{d\phi }\right) ^{2}\right]
\partial_{\mu }\phi
\partial^{\mu } \phi -\frac{V(\phi)}{[ 1+f(\phi )]^{2}}\right\}\
.
\label{2.3}
\end{equation}
The kinetic function can be written as $W(\phi )=\frac{ \omega
(\phi )}{1+f(\phi)}
+\frac{3}{\kappa ^{2}(1+f(\phi ))^{2}}
\left( \frac{df(\phi )}{d\phi }\right) ^{2}$, and the extra term in the
scalar potential can be absorbed by defining the new potential
$U(\phi )=\frac{ V(\phi )}{ \left[ 1+f(\phi ) \right]^2}$, so
that we recover
the action~(\ref{1.1}) in the
Einstein frame, namely
\begin{equation}
S=\int dx^{4} \sqrt{-\widetilde{g}}\left(
\frac{\widetilde{R}}{ \kappa ^{2}}-
\frac{1}{2} \, W( \phi ) \, \partial _{\mu }\phi \partial ^{\mu}
\phi -U(\phi)\right)\ .
\label{2.4}
\end{equation}
We assume that the metric is FRW and spatially flat in this
frame
\begin{equation}
d\widetilde{s}^{2}=-d\widetilde{t}^{2}
+\widetilde{a}^{2}(\widetilde{t})\sum_{i}dx_{i}^{2}\ ,
\label{2.5a}
\end{equation}
then, the equations of motion in this frame are given by
\begin{eqnarray}
&& \widetilde{H}^{2}=\frac{\kappa ^{2}}{6}\rho _{\phi } \;,\\
&&\nonumber \\
&& \dot{\widetilde{H}}=-\frac{\kappa ^{2}}{{4}}\left( \rho
_{\phi } + p_{\phi}\right)\ , \label{2.6}\\
&&\nonumber \\
&& \frac{d^2 \phi}{d\tilde{t}^2} +3\tilde{H} \,
\frac{d\phi}{d\tilde{t}} +\frac{1}{2W(\phi)}\left[ W'(\phi)
\left( \frac{d\phi}{d\tilde{t}}\right)^2 +2U'(\phi) \right] =0
\;,
\end{eqnarray}
where $\rho _{\phi }=\frac{1}{2}W(\phi ){\dot \phi}^{2} +
U(\phi )$, $p_{\phi }=\frac{1}{2}W(\phi ){\dot \phi}^{2} -
U(\phi )$, and
the Hubble parameter is $\widetilde{H}
\equiv \frac{1}{\widetilde{a}}
\frac{d\widetilde{a}}{d\widetilde{t}}$. Then,
\be
W(\phi )\dot{\phi}^{2} =-4\dot{\widetilde{H}}\ ,\quad \quad
U(\phi )=6\widetilde{H}^{2}+2\dot{\widetilde{H}}. \label{2.7}
\ee
Note that $ \dot{\widetilde{H}} >0$ is equivalent to $W<0$;
superacceleration is due to the ``wrong'' (negative) sign of the
kinetic energy, which is the distinctive feature of a phantom
field. The scalar field could be redefined to eliminate
the factor $W(\phi)$, but this would not correct the sign of the
kinetic energy.

If we choose $W(\phi)$ and $U(\phi)$ as $ \omega(\phi)$ and
$V(\phi)$ in~(\ref{1.12}),
\be
W (\phi ) = -\frac{2}{\kappa ^{2}}g^{\prime }(\phi )\ ,\quad
U(\phi ) = \frac{1}{\kappa ^{2}} \left[ {3 g(\phi )}^{2}+ g'
(\phi ) \right]\ ,
\label{ST4}
\ee
by using a function $g(\phi)$ instead of $f(\phi)$
in~(\ref{1.12}), we find a solution as in
(\ref{1.13}),
\begin{equation}
\phi =\widetilde{t}\ , \quad \widetilde{H}(\widetilde{t})=g(\widetilde{t})\ .
\label{2.8}
\end{equation}
In (\ref{ST4}) and hereafter in this section,  we have dropped
the matter contribution for simplicity.

We consider the de Sitter solution in this frame,
\begin{equation}
\widetilde{H}=\widetilde{H}_{0}=\mbox{const.\ } \to \
\widetilde{a}( \widetilde{t})
=\widetilde{a}_{0}e^{\widetilde{H}_{0}\widetilde{t}_{{}}} \;.
\label{2.9}
\end{equation}
We will see below that accelerated expansion can be obtained
in the original
frame corresponding to the Einstein frame~(\ref{2.5a}) with
the solution~(\ref{2.9}),
by choosing an appropriate function $f(\phi )$. From (\ref{2.9})
and
the definition
of $W(\phi )$ and $U(\phi )$, we have
\be
W(\phi )=0 \ \to \ \omega (\phi )=-\frac{3}{\left[ 1+f(\phi)
\right] \kappa^{2}}
\left[ \frac{df(\phi )}{d\phi }\right]^{2}\ ,\quad
U(\phi )=\frac{6}{\kappa ^{2}}\widetilde{H}_{0}^{2}\ \to \
V(\phi )=\frac{6}{\kappa ^{2}}\widetilde{H}_{0}^{2}[1+f(\phi )]^{2}\ . \label{2.10}
\ee
Thus, the scalar field has a non-canonical kinetic term in
the original frame, while in the Einstein frame the latter can
be
positive, depending on $W(\phi)$. The correspondence between
conformal frames can be made explicit through the conformal
transformation~(\ref{2.2}). Assuming a spatially flat FRW metric
in the
original frame,
\begin{equation}
ds^{2}=-dt^{2}+a^{2}(t)\sum_{i=1}^3 dx_{i}^{2}\ , \label{2.10a}
\end{equation}
then, the relation between the time coordinate and the scale
parameter in these frames is given by
\be
t=\int \frac{d\widetilde{t}}{([1+f(\widetilde{t})]^{1/2}}\ ,
\quad a(t)=[1+f(\widetilde{t})]^{-1/2} \,
\widetilde{a}(\widetilde{t})\ .
\label{2.11}
\ee
Now let us discuss the late-time acceleration in the model under
discussion.

As an example, we consider the coupling function
between the scalar field and the Ricci scalar
\begin{equation}
f(\phi )=\frac{1-\alpha \phi }{\alpha \phi }\ , \label{2.12}
\end{equation}
where $\alpha $ is a constant. Then, from~(\ref{2.10}), the
kinetic function $ \omega(\phi)$ and the
potential $V(\phi)$ are
\be
\omega(\phi)=-\frac{3}{\kappa^{2}\alpha^{2}}
\, \frac{1}{\phi^{3}}\ ,
\quad \quad
V(\phi)=\frac{6\widetilde{H}_{0}}{\kappa^{2}
\alpha^{2}}\frac{1}{\phi^{2}}\ ,
\label{2.13}
\ee
respectively. The solution for the current example is found to
be
\be
\phi (t)=\widetilde{t} =\frac{1}{\alpha }\left(
\frac{3\alpha}{2} \, t\right)^{2/3}\ , \quad \quad
a(t)=\widetilde{a}_{0}\left( \frac{3\alpha }{2} \,
t\right)^{1/3}
\exp \left[ \frac{\widetilde{H}_{0}^{{}}}{\alpha } \left(
\frac{3\alpha}{2} \, t\right)^{2/3}\right]\ .
\label{2.14}
\ee
We now calculate the acceleration parameter to study the
behavior of the scalar parameter in the original frame,
\begin{equation}
\frac{\ddot a}{a}=-\frac{2}{9}\frac{1}{t^{2}}+\widetilde{H}_{0}
\left( \frac{2}{3\alpha }\right)^{1/3}\left[ \frac{1}{t^{4/3}}
+ \widetilde{H}_{0}\left( \frac{2}{3\alpha }\right)^{1/3}\frac{1}{t^{2/3}}\right]\ .
\label{2.15}
\end{equation}
We observe that for small values of $t$ the acceleration is
negative; after that we get accelerated expansion for large
$t$; finally, the universe ends with zero acceleration as $t\to
\infty $. Thus, late time accelerated expansion is reproduced by
the action~(\ref{2.1}) with the function $f(\phi )$ given by
Eq.~(\ref{2.12}).

\subsection{Chamaleon mechanism}
The kind of theories described by the above action (\ref{2.1}) have problems when matter is included and one perform a conformal transformation and the Einstein frame is recovered, then the local gravity tests may be violated by a fith force that appeared on a test particle and the violation of the Equivalence Principle is presented. This kind of problems are well constrained by the experiments to a certain value of the coupling parameter as it is pointed below. Recently, a very interesting idea originally proposed in Ref. \cite{chamaleon} avoids the constrains from local gravity tests in such a way that the effects of the scalar field are negleible at small scales but it adquires an important role for large scales, whose effects  may produce the current acceleration of the Universe \\
Let us start by rewriting the action in the Einstein frame (\ref{2.3}) in a similar form as the original Brans-Dicke action by redefining the scalar field $\phi$ and rewriting the kinetic term $\omega(\phi)$ in terms of the coupling $(1+f(\phi))$, then the action is given by:
\be
S_E=\int d^4x\sqrt{-\widetilde{g}}\left[\frac{\widetilde{R}}{2\kappa^2}-\frac{1}{2×}\partial_\mu\sigma\partial^\mu\sigma-U(\sigma)+\e^{4\beta\sigma}\widetilde{L_m} \right]\ , 
\label{3.1}
\ee  
where:
\be
\e^{-2\beta\sigma}=1+f(\phi)\ ,
\label{3.2}
\ee
here $\beta$ is a constant. As it is seen in the action (\ref{3.1}), the matter density lagrangian couples to the scalar field $\sigma$, such that a massive test particle will be under a fith force, and the equation of motion will be:
\be
\ddot{x}^\mu +\widetilde{\Gamma}^\mu_{\lambda\nu}\dot{x}^\lambda \dot{x}^\nu=-\beta\partial^\mu\sigma\ ,
\label{3.3}
\ee
where $x^\mu$ represents the four-vector describing the path of a test particle moving in the metric $\widetilde{g}^(\mu\nu)$, and $\widetilde{\Gamma}^\mu_{\nu\lambda}$ are the Christoffel symbols for the metric $\widetilde{g}^{\mu\nu}$. From the equation (\ref{3.3}), the scalar field $\sigma$ can be seen as a potential from a force given by:
\be
F_{\sigma}=-M\beta\partial^\mu\sigma\ ,
\label{3.4}
\ee
where $M$ is the mass of a test particle. Then, this kind of theories reproduces a fith force which it has been tested by the experiments to a limit $\beta<1.6 \times 10^{-3}$ (Ref. \cite{5force}). The aim of the chamaleon mechanism is that it makes the fith force negleible for small scales passing the local tests. Such mechanism works in the following way, by varying the action (\ref{3.1}) with respect the scalar field $\sigma$, the equation of motion for the scalar field  is obtained:
\be
\bigtriangledown^2\sigma=U_{,\sigma}-\beta \e^{4\beta\sigma}\widetilde{g}^{\mu\nu}\widetilde{T}_{\mu\nu}\ , 
\label{3.5}
\ee
where the energy-momentum tensor is given by $\widetilde{T}^{\mu\nu}=\frac{2}{\sqrt{-\widetilde{g}}×}\frac{\partial L_m}{\partial\widetilde{g}^{\mu\nu}×}$. For simlicity we restrict to dust matter $\widetilde{g}^{\mu\nu}\widetilde{T}_{\mu\nu}=-\widetilde{\rho}_m$, where the energy density may be written in terms of the conformal transformation (\ref{3.2}) as $\rho_m=\widetilde{\rho}_m\e^{3\beta\sigma}$. Then, The equation for the scalar field (\ref{3.5}) is written in the following way:
\be
\bigtriangledown^2\sigma=U_{,\sigma}+\beta\rho_m \e^{\beta\sigma}\ , 
\label{3.6}
\ee
here it is showed that the dynamics of the scalar field depends on the matter energy density. We may write the right side of the equation (\ref{3.6}) as an effective potential $U_{eff}=U(\sigma)+\rho_m\e^{\beta\sigma}$. Then, the behaviour of the scalar field will depend on the effective potential, and the solutions for the equation (\ref{3.6}) are given by studying $U_{eff}$. By imposing to the scalar potential $U(\sigma)$ to be a monotonic decreasing function, the effective potential will have a minimum that will govern the solution for the scalar field (for more details see \cite{chamaleon}), this minimum is given by:
\be
 U_{,\sigma}(\sigma_{min})+\beta\rho_m \e^{\beta\sigma_{min}}=0\ ,
\label{3.7}
\ee
which depends on the local matter density. At this minimum, the scalar field mass will be given by:
\be
m^2_{\sigma}=U_{,\sigma\sigma}(\sigma_{min})+\beta^2\rho_m \e^{\beta\sigma_{min}}\ .
\label{3.8}
\ee
Then, because of the characteristics of the scalar potential $U(\sigma)$, larger values of the local density $\rho_m$ corresponds to small values of $\sigma_{min}$ and large values of $m_{\sigma}$, so it is possible  for sufficiently large values of the scalar field mass to avoid Equivalence Principle violations and fith forces on the Earth. As the energy density $\rho_m$ becomes smaller, the scalar field mass $m_{\sigma}$ decreases and $\sigma_{min}$ increases, such that at large scales (when $\rho\sim H^2_0$), the effects of the scalar field become detected, where the accelerated expansion of the Universe may be a possible effect. Then, one may restrict the original scalar potential $V(\phi)$ and the coupling $(1+f(\phi))$ trough the mechanism showed above. The effective potential $U_{eff}$ is written in terms of $\phi$ by the equation (\ref{3.2}) as:
\be
U_{eff}(\phi)= U(\phi)+\frac{\rho_m}{(1+f(\phi))^{1/2}×}\ ,
\label{3.9}
\ee
where $U(\phi)=V(\phi)/(1+f(\phi))$. Then, by giving a function $f(\phi)$ and a scalar potential $V(\phi)$, one may construct using the conditions described above on the mass $m_\sigma$, a cosmological model that reproduces the current accelerated expansion and at the same time, avoid the local test of gravity.

\section{Conclusions}
Scalar-tensor theories have been studied along the article, where quintessence/phantom models and Brans-Dicke-like theory have been presented and cosmological evolution have been reproduced. There are some problems in this kind of theories, as for example in the models that try to reproduce the whole expansion history, where  the grateful exit from inflation such that the large scale structure should be well studied. In spite of the possible succesful of scalar-tensor theories, the cosmological constant problem still remain as one of the deepest problems in theoretical physics, so it may be researched depper. Also, the coincidence problem that may be resolved by scalar-tensor theory, have not yet a natural explanation. In other hand, the interesting chamaleon mechanism (see \cite{chamaleon}) supposes one way to avoid local gravity tests for the non-minimally scalar-tensor theories, and then, it may be a good candidate to dark energy, although some more relativistic models should be constructed. 
\begin{acknowledgments}
I thank  Emilio Elizalde and Sergei D. Odintsov for  the discussions and collaborations on the current article. This work was supported  by MEC (Spain), project FIS2006-02842, and in part by project PIE2007-50/023.

\end{acknowledgments}

\end{document}